\documentstyle[prd,aps,preprint,amsfonts]{revtex}
\begin{document}
\tighten
\title{Thermalization of a Quark--Gluon Plasma
       \thanks{Work supported by GSI.
       GSI-Preprint 96-57, subm. to Phys.Rev. D}}
\author{P.A.Henning${}^{a,b}$,
        M.Blasone${}^{a,c}$, R.Fauser${}^{a}$ and P.Zhuang${}^{a}$}
\address{${}^{a}$Theoretical Physics,
        Gesellschaft f\"ur Schwerionenforschung GSI\\
        Planckstr. 1, D-64291 Darmstadt, Germany}
\address{${}^{b}$Institut f\"ur Kernphysik, TH Darmstadt\\
         Schlo\ss gartenstra\ss e 9, D-64289 Darmstadt, Germany}
\address{${}^{c}$Dipartimento di Fisica dell'Universit\`a       
         and INFN, Gruppo Collegato,\\ 
         I-84100 Salerno, Italy} 
\date{30. December 1996}
\maketitle
\begin{abstract}
The thermalization time for a Quark-Gluon-Plasma is estimated 
from a quantum transport model beyond the quasi-particle approach
(or kinetic gas theory). While our ansatz is crude concerning the 
properties of ''real'' quarks and gluons, it nevertheless takes very serious
the basic principles of quantum field theory for non-equilibrium states.
It is found, that the thermalization time obtained from quantum transport
theory is substantially longer than from kinetic theory. In our view
this casts some doubts on scenarios which a priori assume a thermalized
quark gluon plasma.
\end{abstract}
\section{Introduction}
In the past decade the effort invested in ultrarelativistic heavy ion 
collisions (URHIC) has grown considerably \cite{QM95}. The general
hope is, that at some time in the near future one may be able to observe an
excursion of strongly interacting matter from the state 
of hadrons before the collision into the phase of a quark--gluon plasma
(QGP). Consequently, the discussion of possible signals from such
a shortlived state is quite vivid: Weakly interacting probes
like photons or lepton pairs, as well as strongly interacting
signals like those presented by quark flavors of higher mass have
been proposed. Similar to most of these
investigations is the assumption of a {\em thermalized\/} plasma
phase, followed by the calculation of the time
evolution along one or the other line of physical reasoning.

With the present paper we address the question, whether 
such a thermalized phase is reached at all. We limit ourselves to
the physical scenario that one may reach in future URHIC: A sea of gluons,
initially at low temperature, is heated to a very high temperature
over a short time. In this {\em hot glue\/}, 
quark-antiquark pairs are popping up -- until at the very end a thermal
equilibrium in the sense of a degenerate plasma is reached. Of
great interest also for the future experiments 
is the time scale for this equilibration.

If one considers quarks and gluons in a standard (Boltzmann-like) transport
theory, this time scale is obtained as $\approx$ 1 fm/c -- short enough
to reach a pseudo-equilibrium in the time span available in heavy ion
collisions. However, we have serious doubts that
the requirements for the applicability of 
kinetic gas theory are fulfilled in a QGP: The thermal
scattering of constituents occurs so frequently, that
subsequent collisions overlap quantum mechanically.

In other words, we expect that in a QGP the {\em off-shell\/}
propagation of particles plays an important role and therefore the
quasi-particle approximation is not applicable. Formally, 
we describe this by including a nontrivial spectral function 
of the constituents, i.e., in the description we go beyond
the quasi-particle approximation. However, the spectral broadening of
particles in a hot plasma, although described on the same footing
as a gas of resonances, does not imply that our particles may decay.
Their continuous mass spectrum merely represents the thermal
scattering by the other components of the system.

For the limited purpose of the present paper, we furthermore make
some physically motivated assumptions:
\begin{enumerate}
\item We assume, that the self energy function for the
quarks is dominated by gluonic contributions. This is justified because the
quark-quark scattering cross section is much smaller than the quark-gluon
cross section.
\item The gluon background is dominated by external conditions, 
i.e., we neglect the back-reaction of quarks on the gluon distribution.
\item The external conditions determining the gluon field are changing
in a short time interval, and the system is translationally invariant in 
3-dimensional coordinate space.
\end{enumerate}
For these assumptions, we also have a practical reason: They allow
for a clean separation of various aspects of the quantum transport problem, 
whereas this separation is difficult (if not impossible)
when considering more realistic systems.
While our primary motivation to go beyond the
quasi-particle picture therefore is a physical one, we may also
adopt a mathematically rigorous stance. The Narnhofer-Thirring theorem
\cite{NRT83} states, that interacting systems at finite
temperature {\em cannot\/} be described by particles with a sharp
dispersion law. Ignoring this mathematical fact one finds as an echo
serious infrared divergences in high temperature quantum chromodynamics
(QCD). Consequently, these unphysical singularities are naturally
removed within the approach of finite temperature field theory with
continuous mass spectrum \cite{L88}.

The effect of off-shell propagation on relaxation processes 
was first investigated by Danielewicz \cite{D84a}. 
He found a substantial slowdown of the
equilibration, when solving the full quantum transport equation
as compared to the solution of its kinetic approximation. 
In later years several attempts were made to unify transport theory with 
concepts going beyond the quasi-particle picture \cite{M86,MH93}.
The connection between transport theory and quantum field theory with 
continuous mass spectrum was established in refs. \cite{h94rep,h94gl3}. 
A first application of these principles to the QGP relaxation problem 
\cite{R94,hny96} points into the same direction as the results of Danielewicz.
 
The paper is organized as follows: In the next section we give a brief
introduction into the formalism necessary for non-equilibrium
quantum fields. In section III we discuss our approximate spectral
function, followed by a solution of the quantum transport equation
in section IV. In section V we show, how one may derive a transport
equation beyond the Boltzmann-like quasi-particle approximation from
the equations in section III, and we solve it in section VI.
Conclusions are drawn in the final section of the present work.
\section{Matrix-valued Schwinger-Dyson equation}
As has been pointed out by various authors,
the description of dynamical (time dependent) quantum phenomena
in a statistical ensemble necessitates a formalism with a doubled
Hilbert space \cite{D84a,h94rep}.  For our purpose the
relevant content of this formalism is that its two-point Green
functions are 2$\times$2 matrix-valued. We leave it to the reader
to chose either the conventional Schwinger-Keldysh,
or Closed-Time Path (CTP) Green
function formalism \cite{SKF,KB62}, or the technically
simpler method of Thermo Field Dynamics (TFD) \cite{Ubook}.

Within this matrix formulation, we consider
the Schwinger-Dyson equation for the full quark propagator
\begin{equation}\label{k1}
           S =       S_0    +      S_0 \odot \Sigma \odot S
\;.\end{equation}
Here $S_0$ is the free and $S$ the full two-point Green function
of the quark field, $\Sigma$ is the full self energy
and the generalized product of these is to be understood as
a matrix product (thermal and spinor indices) and an
integration (each of the matrices is a function of two space
coordinates):
\begin{equation}\label{ovu}
\Sigma^{ij}_{xy}\odot S^{jk}_{yz} = \sum\limits_j\int\!\!d^4y\,
              \Sigma^{ij}_{xy} S^{jk}_{yz}
\;.\end{equation}
Throughout this paper we use the convention to write
space-time and momentum variables also as
lower indices, e.g. $\Sigma_{xy}\equiv \Sigma(x,y)$.

For the purpose of treating transport equations, we switch to the mixed
(or Wigner) representation of functions depending on two
space-time coordinates: 
\begin{equation} \label{wigdef}
\tilde\Sigma_{XP} = \int\!\!d^4(x-y) \;
  \exp\left({\mathrm i} P_\mu (x-y)^\mu\right)\; \Sigma_{xy}
\;.\end{equation}
The $\tilde{}$-sign will be dropped henceforth. 
The Wigner transform of the convolution $\Sigma\odot G$ as defined 
in (\ref{ovu}) is a nontrivial step: Formally it may be expressed as 
a gradient expansion
\begin{equation}\label{gex}
 \int\!\!d^4(x-y) \;
  \exp\left({\mathrm i} P_\mu (x-y)^\mu\right)\; \Sigma_{xz}\odot G_{zy}
 = \tilde\Sigma_{XP} \,
  \exp\left(-{\mathrm i}
  \stackrel{\leftrightarrow}{\Diamond}\right)\,\tilde{G}_{XP}
\;.\end{equation}
$\Diamond$ is a 2nd order differential operator acting to the left and
to the right side,
\begin{equation}
\stackrel{\leftrightarrow}{\Diamond}    =  
\frac{1}{2}\left(\stackrel{\leftarrow}{\partial_X}
\stackrel{\rightarrow}{\partial_P} 
-\stackrel{\leftarrow}{\partial_P}
\stackrel{\rightarrow}{\partial_X}\right)
\;.\end{equation}
The arrows on  the derivatives display the directions in  which  they act.  
Explicitly, the first-order term is
\begin{equation}\label{dia1}
\Sigma_{XP}\,\stackrel{\leftrightarrow}{\Diamond} G_{XP} = \frac{1}{2}\left(
   \frac{\partial \Sigma_{XP}}{\partial X_\mu}
   \frac{\partial G_{XP}}{\partial P^\mu}
  -\frac{\partial \Sigma_{XP}}{\partial P_\mu}
   \frac{\partial G_{XP}}{\partial X^\mu} \right)
\;,\end{equation}
i.e., this is the Poisson bracket of the two quantities
$G$ and $\Sigma$. 

In the CTP formulation as well as in the $\alpha=1$ parameterization
of TFD \cite{hu92}, the matrix elements of $S$, the time-ordered
propagator $S^{11}$, the anti-time-ordered propagator $S^{22}$ and the
1P-correlations $S^{12}$ and $S^{21}$, are not linearly independent
\begin{equation}\label{sme}
S^{11}+S^{22}-S^{12}-S^{21}=0
\;.\end{equation}
A similar relation holds for the free propagator $S_0$ and with
different signs for the  components
of the self energy:
\begin{equation}\label{sse}
\Sigma^{11}+\Sigma^{22}+\Sigma^{12}+\Sigma^{21}=0
\;.\end{equation}
These relations lead to different representations of the retarded and
advanced propagators and self-energies
\begin{eqnarray}\label{sra}\nonumber
S^R =  S^{11}-S^{12}=S^{21}-S^{22}\;,\;\;\;\; &S^A = S^{11}-S^{21}=S^{12}-S^{22}\\
\Sigma^R = \Sigma^{11}+\Sigma^{12}= -\Sigma^{21}-\Sigma^{22}\;,\;\;\;\;
&\Sigma^A = \Sigma^{11}+\Sigma^{21}=- \Sigma^{12}-\Sigma^{22} 
\;.\end{eqnarray}
Due to the above linear relations, 
the four components of the Schwinger-Dyson equation are
not independent, the matrix equation can be simplified by a linear 
transformation. This linear transformation, which one may
conveniently express as a matrix transformation \cite{RS86,hu92},
has a physical interpretation only in the TFD formalism,
see ref. \cite{h94rep}. The transformation matrices ${\cal B}$ are 
\begin{equation}\label{lc}
{\cal B}(n) =
\left(\array{cc}(1\! -\! n) &\;\; -n\\
                1     & 1\endarray\right),\enspace 
({\cal B}(n))^{-1} =
\left(\array{lr}1 &\; n\\
                -1     & (1\!-\!n)\endarray\right)
\;,\end{equation}
depending on one parameter only. Due to the linear relations 
(\ref{sme}) and (\ref{sse}), we obtain for {\em any} value of this parameter
\begin{equation}\label{qptp}
  {\cal B}(n)\,\tau_3\, S\,({\cal B}(n))^{-1}
  = \left({\array{cc}   S^R & \;\;nS^{21}\!+\!(1\!-\!n)S^{12} \\
               0        &  S^A \endarray}\right)
\;,\end{equation}
with $\tau_3 = \mbox{diag}(1,-1)$.

For equilibrium systems the off-diagonal elements of the original matrix valued
propagator fulfill the Kubo-Martin-Schwinger (KMS) condition \cite{KMS}
\begin{equation}\label{kmf}
\left(1 - n_{F}(p_0)\right)S^{12}(p_0,\bbox{p}) +
 n_{F}(p_0) S^{21}(p_0,\bbox{p}) = 0
\,,\end{equation}  
where $n_{F}(E)$ is the Fermi-Dirac equilibrium distribution function at 
temperature $T$, 
\begin{equation}\label{nfdf}
n_{F}(E) = \frac{1}{\displaystyle\mbox{e}^{ \beta (E-\mu)}+1}
\;.\end{equation}
It is obvious, that therefore in equilibrium systems the 
transformation (\ref{qptp}) with $n=n_{F}$ 
results in complete diagonalization of the matrix $S$ \cite{hu92,h94rep}. 
The same distribution function may also be used to diagonalize the matrix
valued self energy function, because it fulfills a relation similar to
(\ref{kmf}). 

In the next step the above condition is generalized to
\begin{equation}\label{kmfg}
\left(1 - N_{XP}\right)S^{12}_{XP} +
 N_{XP} S^{21}_{XP} = 0
\,,\end{equation}  
which {\em defines} the non-equilibrium distribution function $N_{XP}$
depending on coordinates and momenta. In other words, we choose as
non-equilibrium distribution function the parameter which diagonalizes
the non-equilibrium propagator under the transformation (\ref{qptp}):
\begin{equation}\label{sdia}
{\cal B}(N_{XP})\,\tau_3\,S_{XP}\,({\cal B}(N_{XP}))^{-1}=
\left({\array{cc} S^R_{XP} &  0 \\
0  & S^A_{XP} \endarray}\right)
\;.\end{equation}
Generally however, this parameter $N_{XP}$ does {\em not} diagonalize the 
non-equilibrium self energy function. For this purpose one has to introduce
another parameter $N^{\Sigma}_{XP}$, and the transformation of the matrix
valued self energy function then yields
\begin{equation}\label{sigdia}
  {\cal B}(N_{XP})\,\Sigma_{XP}\,\tau_3\,({\cal B}(N_{XP}))^{-1}
  = \left({\array{lr}   \Sigma^R_{XP} & 
\;\;2\pi {\mathrm i} (N^\Sigma_{XP}\!-\!N_{XP})\Gamma_{XP} \\
           0            &  \Sigma^A_{XP} \endarray}\right)
\;,
\end{equation}
where $\Gamma$ is the (common) imaginary part of retarded and advanced 
self energy.  From this splitting follows that the off-diagonal element 
in Eq. (\ref{sigdia}) vanishes for
equilibrium states, because then $N_{XP}=N^{\Sigma}_{XP}=n_{F}(p_{0})$.
This we may use as an indication that the off-diagonal term of (\ref{sigdia}) 
is the "collision term" of a non-equilibrium system. 
  
The splitting into real and imaginary part is useful for self energy as well
as propagator, 
\begin{eqnarray}\label{split}\nonumber
S^{R,A}_{XP} &=& G_{XP} \mp {\mathrm i} \pi {\cal A}_{XP} \\
\Sigma^{R,A}_{XP} &=&
  \mbox{Re}\Sigma_{XP} \mp {\mathrm i}\pi \Gamma_{XP}
\;.\end{eqnarray}
Each of these functions is assumed to be real, and they 
are spinor-valued. The part of the 
self energy function which is local in space and time, is in our 
notation included in $\mbox{Re}\Sigma_{XP}$. For a relativistic model this
is the Hartree part of $\Sigma$. 

From eqns. (\ref{sdia}) and (\ref{sigdia}) follows, that
the diagonal elements of the matrix transformed Schwinger-Dyson equation 
are {\em retarded\/} and {\em advanced\/} Schwinger-Dyson equation.
The off-diagonal element is a {\em transport equation}, because it determines
the non-equilibrium distribution function. As said above, 
the driving force of the
transport part is the deviation of the non-equilibrium
distribution function from the corresponding equilibrium distribution
function to which the whole system tends. 

A special virtue of this formulation is the fact, that with the KMS condition
(\ref{kmf}) we have already shown that this non-equilibrium distribution
function reaches a Fermi-Dirac distribution in case we are considering an
equilibrium system.

We now consider the equations obtained by action of Dirac
differential operators (= {\em inverse free propagators\/}) on the
matrix-transformed Schwinger-Dyson equation. They act on the free
retarded and advanced propagator as
\begin{equation}\label{smix}
\widehat{S}_0^{-1}\,S^{A,R}_{0,XP} =
\left(P^\mu\gamma_\mu -M +\frac{{\mathrm i}}{2}\gamma_\mu \partial_X^\mu\right)
\,S^{A,R}_{0,XP}=1
\;.\end{equation}
Acting with the inverse free propagator on the full Schwinger-Dyson
equation, we obtain
\begin{equation}
\widehat{S}_0^{-1}\,S^{A,R}_{XP}=
1+\Sigma^{A,R}_{XP}\,\exp\left(-{\mathrm i}
 \stackrel{\leftrightarrow}{\Diamond}\right)\,S^{A,R}_{XP}  
\;.\end{equation}
Adding and subtracting the equations for the retarded and advanced
propagators, we obtain equations of motion for the spectral function
and the real part of the retarded and advanced propagator
\begin{eqnarray}\label{diae}
2{\mathrm i}\pi \widehat{S}_0^{-1}\,{\cal A}^{A,R}_{XP} &=&
\Sigma^{A}_{XP}\,\exp\left(-{\mathrm i}
  \stackrel{\leftrightarrow}{\Diamond}\right)\,S^{A}_{XP}  
-\Sigma^{R}_{XP}\,\exp\left(-{\mathrm i}
  \stackrel{\leftrightarrow}{\Diamond}\right)\,S^{R}_{XP} 
\;,\label{eqsp}\\
\widehat{S}_0^{-1}\,G^{A,R}_{XP} &=&
1+\frac{1}{2}\left(\Sigma^{R}_{XP}\,\exp\left(-{\mathrm i}
  \stackrel{\leftrightarrow}{\Diamond}\right)\,S^{R}_{XP}  
+\Sigma^{A}_{XP}\,\exp\left(-{\mathrm i}
  \stackrel{\leftrightarrow}{\Diamond}\right)\,S^{A}_{XP} \right)
\end{eqnarray}
for the diagonal components of the transformed Schwinger-Dyson equation.
The solution in zero order $\Diamond$ leads to the solution for a 
equilibrium system. The equations in first order of the gradient approxiamtion
give a transport equation for the spectral function which is in equilibrium 
solved by the solution in zero order $\Diamond$.

A very important fact about these equations has to be emphasized 
\cite{h94gl3}: In general, they do not admit a $\delta$-function solution
for the spectral function ${\cal A}_{XP}$ even in zero order of the
gradient expansion. This has led to much confusion in papers deriving
transport equations from the Schwinger-Dyson equation. We therefore
state very clearly what is obvious for mathematical as well as
physical reasons: There is not such thing as a mass shell constraint in 
quantum transport theory ! To insert such a constraint into any model by hand
must be considered dangerous unless it is tested against the full calculation.

The off-diagonal component of the transformed Schwinger-Dyson equation
reads, after acting on it with the inverse free propagator,
\begin{equation}\label{k5}
\widehat{S}^{-1}_0   S^K_{XP}     = 
\Sigma^R_{XP} \,\exp\left(-{\mathrm i}\stackrel{\leftrightarrow}{\Diamond}\right)\, S^K_{XP}
- \Sigma^K_{XP}\,\exp\left(-{\mathrm i}\stackrel{\leftrightarrow}{\Diamond}\right)\, S^A_{XP}
\;\end{equation}
(see \cite[pp.307]{h94rep} for details). 
A similar equation holds for the propagator $S^{12}$
\begin{equation}
\widehat{S}^{-1}_0   S^{12}_{XP}     =  
\Sigma^R_{XP}\,\exp\left(-{\mathrm i}
 \stackrel{\leftrightarrow}{\Diamond}\right)\, S^{12}_{XP} + \Sigma^{12}_{XP}
\,\exp\left(-{\mathrm i}\stackrel{\leftrightarrow}{\Diamond}\right)\, 
S^A_{XP}
\;\end{equation}
Using the equation for the spectral function (eq. (\ref{eqsp})) and 
$S^{12}_{XP}=2{\mathrm i}\pi N_{XP} {\cal A}_{XP}$,
we obtain a differential equation for $N_{XP}$
\begin{eqnarray}\label{trae}
- \pi \gamma_\mu{\cal
A}_{XP}\partial_X^\mu N_{XP}
&= &  \Sigma^R_{XP}\,\exp\left(-{\mathrm i}
   \stackrel{\leftrightarrow}{\Diamond}\right)\, S^{12}_{XP} 
+ \Sigma^{12}_{XP}\,\exp\left(-{\mathrm i}
  \stackrel{\leftrightarrow}{\Diamond}\right)\,
S^A_{XP} \nonumber\\
&&+N_{XP}\Big(\Sigma^R_{XP}\,\exp\left(-{\mathrm i}
  \stackrel{\leftrightarrow}{\Diamond}\right)\, S^R_{XP}
-\Sigma^A_{XP}\,\exp\left(-{\mathrm i}
  \stackrel{\leftrightarrow}{\Diamond}\right)\, S^A_{XP}\Big)
\;.\end{eqnarray}
The transport equations are usually derived  by the terms up to first order 
$\Diamond$, but it is visible that transport equations contain in principle 
arbitrary orders of the operator $\Diamond$. Below for a certain approximation 
the transport equation will be solved up to all orders in the gradient 
expansion. 

\section{Ansatz for self energy and spectral functions}
As we have seen above, the matrix valued propagator has only
three independent components, two of which are furthermore complex conjugate.
We have also shown how one may  use the KMS condition to eliminate the 
off-diagonal component of this propagator in favor of the distribution 
function parameter. 

We now tackle the remaining two pieces, i.e., we deal with retarded and
advanced propagator. In coordinate space they are zero in the forward
resp. backward lightcone (outside, both are identically zero),
\begin{equation}\label{raff2}
S^{R,A}_{xy} = \mp 2\pi{\mathrm i}\Theta\left(\pm(x_0-y_0)\right)
        {\cal A}_{xy}
\;.\end{equation}
Consequently, their Wigner
representations are analytical functions in
the upper resp. lower complex energy halfplane.
Hence, even for non-equilibrium states we may write in the
mixed (or Wigner) representation
\begin{equation}\label{rapf}
S^{R,A}(E,\bbox{p},X)  = \mbox{G}_{XP} \mp \pi {\mathrm i} {\cal A}_{XP} =
  \int\limits_{-\infty}^{\infty}\!\!dE^\prime\;
  \frac{ {\cal A}(E^\prime,\bbox{p},X) }{E-E^\prime\pm{\mathrm i}\epsilon}
\;.\end{equation}
Without elaboration at this point we emphasize,
that one may {\em not\/} calculate the real part of the quark propagator
by a method different from eq. (\ref{rapf}), e.g. perturbatively.
In a relativistic system, doing so may lead to 
unphysical poles in the retarded propagator, on the wrong side of
the real energy axis -- the retarded propagator then does {\em not\/}
fulfill eq. (\ref{raff2}). 

The principal difference between equilibrium and non-equilibrium states at
this point are the properties of the function 
\begin{equation}\label{gsp}
{\cal A}(E,\bbox{p}) = \mp\frac{1}{\pi}\,\mbox{Im}(S^{R,A}(E,
\bbox{p}))
  = \frac{1}{2\pi{\mathrm i}}\left(S^A(E,\bbox{p}) - S^R(E,
\bbox{p})\right)
\;.\end{equation}
In equilibrium states it has {\em spectral} properties, i.e., it 
is normalized and {\em positive
semidefinite}. In non-equilibrium states positivity is not guaranteed,
we can only use the term "spectral function" in a generalized fashion.
Normalization of the $\gamma^0$-component of ${\cal A}$ according to
\begin{equation}\label{norm}
\int\limits_0^\infty\!dE\,\mbox{Tr}\left[ \gamma^0\,{\cal A}(E,
\bbox{p})\right]
 = 2
\;\end{equation}
is guaranteed also in non-equilibrium states: It
is a consequence of the canonical anti-commutation relations for the fields. 

For equilibrium states we may now combine the facts into a very compact
notation for the matrix valued propagator, using the
Bogoliubov matrix defined in (\ref{lc})  \cite{hu92,h94rep}
\begin{equation}\label{fsk1}
S^{(ab)}(p_0,\bbox{p}) =
   \int\limits_{-\infty}^\infty\!\!dE\,
        \tau_3\, \big({\cal B}(n_{F}(E))\big)^{-1}\;
   \left(\!{\array{cc}
         {\displaystyle \frac{{\cal A}(E,\bbox{p})}{p_0-E+{\mathrm
   i}\epsilon}}
 & 0 \\  
 0 &\,\,{\displaystyle \frac{{\cal A}(E,\bbox{p})}{p_0-E-{\mathrm i}\epsilon}}
\endarray}\right)\;
 {\cal B}(n_{F}(E))
\;.\end{equation}
For free fermions of mass $m$ the spectral function is
\begin{equation}\label{ff}
{\cal A}(E,\bbox{p}) \longrightarrow
        \left( E\gamma^0 + \bbox{p}\bbox{\gamma} + m\right)\,
        \mbox{sign}(E)\,\delta(E^2-\bbox{p}^2 - m^2)
\;.\end{equation}
Instead of attempting a fully self-consistent numerical calculation of
the spectral function for non-equilibrium states, 
we make another approximation.

It is assumed, that the spectral function of the interacting system 
does not differ too much from the spectral function of
quasi-particles. In particular, we make the ansatz:
\begin{eqnarray} \nonumber
{\cal A}(E,\bbox{p},t) &=& \frac{\gamma_t}{\pi} 
\frac{\gamma^0\left(E^2 + \omega_t^2+\gamma_t^2\right) +
      2 E \bbox{\gamma}\bbox{p} + 2 E m_t}{
  \left(E^2 - \omega_t^2-\gamma_t^2\right)^2 + 4 E^2 \gamma_t^2}
\\ \nonumber
 &=&\frac{1}{4\pi{\mathrm i}\omega_t} \left(
\frac{\omega_t
  \gamma^0 + \bbox{p}\bbox{\gamma} + m_t}{E - \omega_t - 
   {\mathrm i} \gamma_t}
-\frac{-\omega_t\gamma^0 + \bbox{p}\bbox{\gamma} + m_t}{
E+\omega_t
 - {\mathrm i} \gamma_t}\right.
\\ \label{sfan}
 &&\hphantom{\frac{1}{4\pi{\mathrm i}\omega_t}} \left.
-\frac{\omega_t\gamma^0 + \bbox{p}\bbox{\gamma} + m_t}{
E - \omega_t
 + {\mathrm i} \gamma_t}
+\frac{-\omega_t\gamma^0 + \bbox{p}\bbox{\gamma} + m_t}{
E+\omega_t
 + {\mathrm i} \gamma_t}
\right)\;,
\end{eqnarray}
where $m_t$ is the effective mass $m_t=m-{\mbox{Re}}\Sigma(\bbox{p}=0)$,
see below.
Hence, we approximate the quark spectral function by two time-dependent 
parameters $\omega_t$ 
and $\gamma_t$, which we may interpret as 
effective quasi-particle energy and effective spectral width and
describe the energy pole as a function of time.
One may argue about the validity 
of this approach, in particular whether not a momentum dependent
spectral width is an absolute necessity for a realistic calculation.

Let us discuss this in three steps. First of all,
in ref.\cite{hq95gam} the same approximation was used to obtain numerical 
results for photon radiation rates from a hot QGP. It was found, that in the
proper temperature and energy regions these results agree very well with
the hard thermal loop approximation scheme of QCD. 

Secondly, we may safely assume that the quarks 
appearing in the hot medium are slow -- hence the properties of
the quark distribution may be safely approximated by those of
quarks at rest. Third, and more important from the aspect of model 
consistency is the fact that the expectation value of the 
anti-commutator of two quark fields is
\begin{equation}
\left\langle \left\{ \Psi(x), \overline{\Psi}(y) \right\} \right\rangle
 = \int\limits_{-\infty}^\infty \!\!dE \,\int\!\frac{d^3\bbox{p}}{
(2\pi)^3}\;{\cal A}\Big(E,\bbox{p},\frac{x^0+y^0}{2}\Big)\,\exp\left(
-{\mathrm i} (E (x^0-y^0) - \bbox{p}(\bbox{x}-\bbox{y}))\right)
\;.\end{equation}
With our ansatz for the spectral function it is easy to show, that
this commutator vanishes for spacelike distances $x-y$ {\em also\/}
in case $\omega_t$ and $\gamma_t$ are time-dependent \cite{hpsb96}.
Conversely, for a general momentum dependence of the parameters 
$\omega$ and $\gamma$ this cannot be guaranteed.

By inspection of eqs. (\ref{diae}) and (\ref{trae}) we find, that 
beyond this only a quark self energy function
is needed for a full determination of the function ${\cal A}_{XP}$.
This self energy function is in general a functional of 
${\cal A}_{XP}$ again -- which then leads to a complicated
set of integro-differential equations for the self consistent determination
of the retarded and advanced propagator. However, for the limited purpose of
the present paper we have formulated simple approximations in the
introductory part of the present paper.

In our equations these ideas are introduced through replacing the
$X,P$ dependence of the self energy completely by a dependence on
$X_{0}\equiv t$, 
\begin{eqnarray}
{\mbox{Re}}\Sigma_{X_0}&={\mbox{Re}}\Sigma_{t}&=\; 
{\mbox{Re}}\Sigma_f\theta(t) +
{\mbox{Re}}\Sigma_i\theta(-t)\,, \nonumber\\ 
\label{sean}
\Gamma_{X_0}&=\Gamma_{t} &=\; 
\widetilde{g}(T)\,\gamma^0  
  \big(T_f \theta(t)+T_i \theta(-t)\big)\, 
\;.\end{eqnarray}
In other terms, we take as the physically most important pieces of the self
energy function the Dirac scalar real part, and the Dirac ''vector'' imaginary
part. The parameters ${\mbox{Re}}\Sigma_i$, ${\mbox{Re}}\Sigma_f$, 
denote the constant quark mass shift before and after time $t=0$,
$T_{i}$ and $T_{f}$ denote the corresponding temperatures. 
$\widetilde{g}(T)$ is some numerical factor given below.

This ansatz  corresponds to instantaneous heating of the gluon
background and describes quarks in the corresponding complex external field.
It is different from a previous treatment off the relaxation problem
\cite{hny96}, where only the imaginary part of the self energy function was
used as input.
\section{Solution of the retarded and advanced equation}
In the following we discuss the solutions for the time-dependent functions 
$\omega_t$ and $\gamma_t$ from our ansatz for the spectral
function, which are obtained when inserting the ansatz for the self-energy.
Indeed, using this self energy it is
possible to solve the r.h.s. of the equation of motions of the
retarded and advanced functions to all orders in the gradient
expansion, see appendix. In particular, the equations 
obtained after taking the trace over the Dirac indices and adding
the real parts of Eq. (\ref{diae}) resp. subtracting the imaginary parts are
\begin{eqnarray}
\frac{1}{4}{\mbox{Re}}{\mbox{Tr}}\Big[
&&\left(\gamma_\mu P^\mu -M +\frac{{\mathrm i}}{2}
\gamma_\mu\partial_X^\mu\right) 
\,\left(S^{R}_{XP}+S^{A}_{XP}\right)\Big]= 2\nonumber\\
&&+\frac{1}{4}{\mbox{Re}}{\mbox{Tr}}
\Big[\Sigma^{R}_{XP}\,\exp\left(-{\mathrm i}
\stackrel{\leftrightarrow}{\Diamond}\right)\,S^{R}_{XP}  
+\Sigma^{A}_{XP}\,\exp\left(-{\mathrm i}
\stackrel{\leftrightarrow}{\Diamond}\right)\,S^{A}_{XP}\Big]\;,
\nonumber\\
\frac{1}{4}{\mbox{Im}}{\mbox{Tr}}\Big[
&&\left(\gamma_\mu P^\mu -M +\frac{{\mathrm i}}{2} \gamma_\mu
\partial_X^\mu \right)
\,\left(S^{R}_{XP}-S^{A}_{XP}\right)\Big]\nonumber\\
&&=\frac{1}{4}{\mbox{Im}}{\mbox{Tr}}
\Big[\Sigma^{R}_{XP}\,\exp\left(-{\mathrm i}
\stackrel{\leftrightarrow}{\Diamond}\right)\,S^{R}_{XP}  
-\Sigma^{A}_{XP}\,\exp\left(-{\mathrm i}
\stackrel{\leftrightarrow}{\Diamond}\right)\,S^{A}_{XP}\Big]
\;.\end{eqnarray}
The complete expressions are given in the appendix.
The choice of the above combinations of taking the real or imaginary
part and adding or subtracting the equations for $S^R$ and $S^A$
ensures that drift term contributions for $S^R$ and $S^A$ resulting
from the derivative $\partial_\mu^X$ on the l.h.s. drop out.
Thus, only the zeroth order terms in the gradient expansion remain on
the l.h.s.. Taking the unphysical limit of vanishing spectral width,
the above equations lead to the mass shell constraint. The other two,
possible combinations pick out the drift term contributions and lead
to  transport equations for the real and imaginary part of $S^R$ and
$S^A$. For a system on non-relativistic particles these transport
equations are fulfilled automatically in gradient expansion if one
inserts the solution for $S^R$ and $S^A$ from the equations of zeroth
order in the gradient expansion \cite{h94rep,BM90}. 
Here, taking into account all orders
of the gradient expansion, these equations lead to two additional
equations which must be fulfilled and are discussed below.

Performing further reductions, one obtains a set of two coupled nonlinear 
equations for $\gamma_t$ and $\omega_t$. 
For a concrete evaluation, the energy
parameter is chosen as $E=\omega_t$. This implies that
the energy parameter is equal to the real part of the
quasi-particle pole, the equations are thus evaluated at the peak of
the spectral function. The result is:
\begin{eqnarray}\label{k9c}
\nonumber
&&\vphantom{\int}
\omega_t^2={\bbox{p}}^2+m_t^2+\frac{\displaystyle
\omega_t\gamma_t(E_1+F_1)-(2\omega_t^2+\gamma_t^2)(E_2+F_2)
}{\displaystyle\omega_t(4\omega_t^2+\gamma_t^2)}\,\nonumber\\
&&\vphantom{\int}
\gamma_t=\pi \widetilde{g} T-\frac{\displaystyle
2 \omega_t(E_1+F_1)-\gamma_t(E_2+F_2)
}{\displaystyle\omega_t(4\omega_t^2+\gamma_t^2)}
\end{eqnarray}
with functions $E_{1,2}$ and $F_{1,2}$ defined as
\begin{eqnarray}\label{k10c}
&&E_{1}=-\frac{m_t
({\mbox{Re}}\Sigma_f-{\mbox{Re}}\Sigma_i)}{4\omega_t}
\Big[ -\theta(t)e^{-2\gamma_t t}
\big(2\omega_t\gamma_t \cos(4\omega_t t)
+\gamma_t^2\sin(4\omega_t t)\big)
\nonumber\\
&&\qquad\qquad +\theta(-t)e^{2\gamma_t t}
\big(2\omega_t\gamma_t \cos(4\omega_t t)
-\gamma_t^2\sin(4\omega_t t)\big)\Big]
\nonumber\;,\\
&&E_{2}=-\frac{m_t
({\mbox{Re}}\Sigma_f-{\mbox{Re}}\Sigma_i)}{4\omega_t}
\Big[ -\theta(t)e^{-2\gamma_t t}
\big(4\omega_t^2+\gamma_t^2-\gamma_t^2 \cos(4\omega_t t)
+2\omega_t\gamma_t\sin(4\omega_t t)\big)
\nonumber\\
&&\qquad\qquad +\theta(-t)e^{2\gamma_t t}
\big(4\omega_t^2+\gamma_t^2-\gamma_t^2 \cos(4\omega_t t)
-2\omega_t\gamma_t\sin(4\omega_t t)\big)
\Big]\nonumber\;,\\
&&F_{1}=-\frac{\pi \widetilde{g} (T_f-T_i)}{4}\Big[ -\theta(t)e^{-2\gamma_t t}
\big(4\omega_t^2+\gamma_t^2+\gamma_t^2 \cos(4\omega_t t)
-2\omega_t\gamma_t\sin(4\omega_t t)\big)
\nonumber\\
&&\qquad\qquad +\theta(-t)e^{2\gamma_t t}
\big(4\omega_t^2+\gamma_t^2+\gamma_t^2 \cos(4\omega_t t)
+2\omega_t\gamma_t\sin(4\omega_t t)\big)\Big]
\nonumber\;,\\
&&F_{2}=-\frac{\pi \widetilde{g} (T_f-T_i)}{4}\Big[ -\theta(t)e^{-2\gamma_t t}
\big(2\omega_t\gamma_t \cos(4\omega_t t)+\gamma_t^2\sin(4\omega_t t)\big)
\nonumber\\
&&\qquad\qquad +\theta(-t)e^{2\gamma_t t}
\big(2\omega_t\gamma_t \cos(4\omega_t t)-\gamma_t^2\sin(4\omega_t t)\big)\Big]
\;.\end{eqnarray}
First we discuss limiting cases of the above equations. For
$t=\pm\infty$ we obtain
\begin{eqnarray}
\nonumber
\omega_{t=\pm\infty}^2&=&{\bbox{p}}^2+(m+{\mbox{Re}}\Sigma_{f/i})^2
= {\bbox{p}}^2 + m_{f/i}^{2}\\
\gamma_{t=\pm\infty}&=&\pi \widetilde{g} T_{f/i} 
\;.\end{eqnarray}
Consequently, our equations have the proper boundary conditions for the 
effective quark mass and the spectral width of the quarks.

Furthermore, one sees that
at $t=0$ the effective width remains smooth and takes the value 
$\gamma_{t=0}= \pi \widetilde{g} (T_f-T_i)/2$, whereas the function $\omega_t$ has a
jump at $t=0$: The limit from the left is
$\omega^2_{t=0_-}={\bbox{p}}^2+m_i(m_i+m_f)/2$ and from the right
is $\omega^2_{t=0_+}={\bbox{p}}^2+m_f (m_i+m_f)/2$. 

In each of these equations it is easy to
distinguish between the contribution of quarks and antiquarks:
If one takes into account only the positive energy pieces of the
spectral function, terms oscillating with frequency $4\omega_t$ drop out.

We now turn to a numerical evaluation of these formal results.
To this end, we first have to specify what we mean by the factor
$\widetilde{g}$ in eq. (\ref{sean}). 
In our view its is not clear to the present
date, how this parameter is related to the strong coupling constant: Some
authors use a quadratic, others a linear dependence on the actual
$g_{S}$. A thorough discussion is carried out in the second paper of ref.
\cite{hq95gam}. Since it was obtained there, that the numerical results
are quite independent of the actual functional structure, we adopt a pragmatic
view in the following and use a simple expression estimated within the hard
thermal loop scheme assuming a two flavor system and quarks with 
zero momentum (${\bbox{p}}^2=0$) \cite{thoma}:
\begin{eqnarray}
&\gamma(T_{f})&=\;\frac{5.63}{12 \pi}\, g^2\, T_{f}=1.88\, \alpha_S\, T_{f}\;,
\nonumber \vphantom{\int}\\
\Leftrightarrow&\widetilde{g}& =\; 0.598\,\alpha_{S}
\label{gti}
\;.\end{eqnarray}
The temperature dependent coupling $g$ follows a parametrization
of the strong coupling constant $\alpha_S={g^2}/{4\pi}$ 
given by Karsch \cite{karsch}:
\begin{equation}
\alpha_S(T)=\frac{6\pi}{(33-2 n_f)\ln(8T/T_c)}\;,
\end{equation}
where $T_c$ is the critical temperature and $n_f$ the number of
flavors. Some values for the coupling $g$ with the corresponding 
temperature and critical temperature are listed in Table 1.
For temperatures less than $T_c$ the spectral width of the quarks is
chosen infinitely small, in agreement with more elaborate
self-consistent calculation within a generalized Nambu -- Jona-Lasinio model 
\cite{hq95gam}.

The effective quark mass for temperatures less than $T_c$ is 
taken as the standard constituent quark mass $m_i$=300
MeV. Above the critical temperature, we take a termal quark mass from
\cite{kajantie}
\begin{equation}
m^2_T=\frac{g^2 T^2}{6}+\frac{g^2  \mu^2}{6 \pi^2}\;.
\end{equation}
The term containing the chemical potential is small in comparison with
the first term and neglected in the calculations. The thermal masses
are shown in Table 2. 

Plugging these parametrizations into our highly nonlinear equations then
requires some effort to solve them numerically.
Let us now comment on the results of this model calculation. 
In Fig. 1 $m_t=\omega_{t}(\bbox{p}=0)$ is depicted as a function of time. 
Furthermore for the special case of $T_f=T_c$ the temperature dependence 
of the solution is  indicated. As shown above, the constitutent quark mass
of 300 MeV is obtained as the boundary value.
Approaching $t=0$ the effective quark mass decreases with time until
it reaches the left-sided limiting  value at $t=0_-$ discussed above. 
Then is jumps down to the right-sided limiting value  at 
$t=0_+$ and decreases further and approaches the final value for 
the quark-gluon plasma. 

Conversely, the spectral width gamma increases with time, from the 
small value in the nuclear environment to a large value in the QGP.
Its value is shown in Fig 2. Note, that due to our ansatz functions
we find a jump in the effective quark mass, whereas 
$\gamma_t$ is smooth at $t=0$.
Both curves exhibit quantum oscillations with the frequency $4\omega_t$,
becoming slower due to the decrease of $\omega_t$ as a function of time.
 
It is furthermore necessary to point out that the effects for $t<0$ are 
no acausal behaviour. They arise in the present formulation 
from the fact that $t$ is the center-of-mass time of a 2-point  
function and not the physical time of the system. Therefore, even for
$t<0$ there are contributions coming from the time after the jump in 
temperature since it is  
possible to choose an adapted relative time. Naturally these contributions
decrease to zero if $t \longrightarrow -\infty$ 
as it can be seen in Fig 1 and 2.

Besides the above two equations we can get two further equations for the 
spectral function by adding the imaginary
parts and by subtracting the real parts of the equations for $S^R$
and $S^A$. 
These equations could in principle be solved if we would not have fixed 
the shape of the spectral function as function of $E$ and ${\bbox{p}}$ by 
the above ansatz. Consequently, in our picture they are fulfilled only
approximately. 
\section{Transport equation}\label{tpe}
We now turn to the primary task of the present paper, i.e., to the evolution
equation for the quark distribution function. 
This transport equation is obtained from eq. (\ref{sse})
by taking the trace over spinor indices 
and separation into real and imaginary part:
\begin{eqnarray}
\label{tr1}
&&{\mbox{Re}}{\mbox{Tr}}\Big[- \pi \gamma_\mu{\cal
A}_{XP}\partial_X^\mu N_{XP}\Big] \nonumber\\
&&={\mbox{Re}}{\mbox{Tr}}\Big[
\Sigma^R_{XP}\,\exp\left(-{\mathrm i}
\stackrel{\leftrightarrow}{\Diamond}\right)\, S^{12}_{XP} 
+ \Sigma^{12}_{XP}\,\exp\left(-{\mathrm i}
\stackrel{\leftrightarrow}{\Diamond}\right)\,S^A_{XP} \nonumber\\
&&+N_{XP}\Big(\Sigma^R_{XP}\,\exp\left(-{\mathrm i}
\stackrel{\leftrightarrow}{\Diamond}\right)\, S^R_{XP}
-(\Sigma^A_{XP}\,\exp\left(-{\mathrm i}
\stackrel{\leftrightarrow}{\Diamond}\Big)\, S^A_{XP}\right)\Big]
\;.\end{eqnarray}
The real part constitutes the desired transport equation, whereas
the imaginary part may be used to check the consistency of e.g. 
the approximation for $N_{XP}^\Sigma$. 

The latter is necessary because, in addition to the approximations made for 
the retarded self energy components, we explicitly need
$\Sigma^{12}$ for the transport equation.
Using a similar ansatz as for  the retarded and advanced self-energy, it is
treated as an external field which exhibits a jump at $t=0$. The
values before and after the jump are given by the boundary conditions 
at $t=\pm\infty$:
\begin{equation}
\Sigma^{12}_{XP}=2 \pi {\mathrm i}\,\widetilde{g} 
  \gamma_0 (\theta(t) N_f T_f +  \theta(-t) N_i T_i)\;, 
\end{equation}
with $N_{f/i}$ being the final resp. the initial quark distribution function,
i.e., the Fermi-Dirac distribution at $T=T_{f/i}$

It is a rather straightforward task to insert this ansatz into the above 
equation, after some elementary manipulations one obtains the
quantum transport equation for the physical system described by our
aproximations as 
\begin{eqnarray}
\partial_t N_t &=&
2\pi \widetilde{g}\,\Big[\theta(t)T_f(N_f-N_t)+\theta(-t)T_i(N_i-N_t)\Big]
\nonumber\vphantom{\int}\\
&&-\pi \widetilde{g}
\Big[T_f(N_f-N_t)-T_i(N_i-N_t)\Big]\theta(t) Z(t)\nonumber\vphantom{\int}\\
Z(t) &=& \frac{e^{-2\gamma_t t}}{2\omega_t^2+\gamma_t^2} 
\Big[4\omega_t^2+\gamma_t^2+\gamma_t^2 \cos(4\omega_t t)
-2\omega_t\gamma_t\sin(4\omega_t t)\Big]\vphantom{\int}
\;,\end{eqnarray}
with $\widetilde{g}$ as in eq. (\ref{gti}).
Note, that the real part of the self-energy does not enter 
the transport equation {\em explicitly}. The spectral width of the quarks is
responsible for the reoccupation in the non-equilibrium situation
after the jump in temperature.

For the interpretation of this result it is useful to study the classical
limit of the above transport equation given by
\begin{equation}
\partial_t N^{cl}_t =
2\pi \widetilde{g}\,
\Big[\theta(t)T_f(N_f-N^{cl}_t)+\theta(-t)T_i(N_i-N^{cl}_t)\Big]
\;.\end{equation}
The classical and the quantum transport equation have two things in common.
First, if the system starts from an equilibrium state, i.e., 
for $t<0$ with $N_t=N_i$ ($N^{cl}_t=N_i$), the occupation
number parameter remains equal to $N_i$ until $t=0$. Furthermore,
in the limit $t\rightarrow+\infty$, $N_t$ ($N^{cl}_{t}$) approaches $N_f$.

Otherwise however, the solutions of the two quations are vastly different.
This difference between the quantum and the classical transport problem
may be characterized by the slope of the solution at
$t=0_+$. The slope of the classical solution at $t=0_+$ is 
$2\pi \widetilde{g} T_f (N_f - N_i)$, while the slope of the full soulution 
is zero. 

This implies, that the reaction of the quantum system to the jump in
temperature does not start immediately as in the classical case, but with
a certain {\em delay time}. In Fig. 3 we plotted the classical and 
the quantum solution for comparison.
As may be inferred from Fig. 3, the quantum delay time decreases with
increasing temeprature. However, the highly nonlinear nature of our coupled 
equations leads to a non-analytical dependence of this delay time on the
temperature.

For our model calculation, this behaviour is examined in more detail 
in Fig. 4, where we plotted values of the time $\tau$ needed
to reach 90 \% of the final occupation number for quarks at rest.
The curve parameter is the critical temperature in the range from
150 to 250 MeV, marking the boundaries of the range of currently 
accepted phase 
transition temperature from nuclear to quark matter.
Values for $\tau$ are given only for $T>T_{c}$.
\section{Conclusion}
To draw conclusions from the present paper, we work backwards starting from
Fig. 4 and 5. The most prominent fact inferred from these figures is 
the lenghtening of the relaxation time for a QGP by a factor $\approx$1.5
due to quantum effects. This result is qualitatively as well as quantitatively
similar to the results obtained for different systems, cf. \cite{D84a}.
Also, the absolute value of the relaxation time is comparable to thermalization
times generally accepted for quark matter \cite{QM95}, but systematically
higher than 1 fm/c. 

In particular: using a phase transition temperature of 160 MeV as obtained in
various realistic calculations, together with a final temperature only
moderately higher than this $T_{c}$, the time $\tau$ 
to reach a 90 \% thermalized 
QGP is $\approx$ 3.6 fm/c. In our view, this time scale is much too long 
to sustain the constant ''hot glue'' scenario. The natural conclusion
therefore would be, that with a moderately ultrarelativistic heavy-ion
collison one might probably see a phase transition to quark matter 
-- but most ceratainly not a {\em thermalized} plasma phase.

The quantum effects we find have a straightforward
physical interpretation: ''collisions'' need a certain time to build up.
However, in our view this straightforward interpretation is very misleading.
We have shown, that the "quantum delay" is due to the spectral width of
the quarks in the hot medium. This spectral width is a manifestly
non-classical effect, to attribute it to "collisions" is sneaking the
mechanistic quasi-particle picture back into the physical description.

That this picture is in profound contradiction to our results is obvious from
the quantum oscillations in Fig. 2 and 3. Such quantum oscillations must be 
taken rather serious, because in a completely
different area of physics they are even established experimentally: They seem
to play an important role in four wave mixing of ultrashort laser pulses
\cite{haug}. Consequently, neither the effective mass, nor the
spectral width parameter are monotonous functions. 

To our knowledge there is no principle which would forbid such oscillations
also in the occupation number parameter $N_{t}$. Indeed, a close look at the
results reveals such oscillations in the derivative of our solution, albeit
not strong enough in the present model to lead to a piecewise diminishing
$N_{t}$.

With the present paper, we have taken the non-equilibrium character of our
quantum system more serious than a realistic description of the interaction.
In particular, we made a physically motivated ansatz for all three independent
elements of the $2\times 2$ matrix valued self energy function. One might
argue, that in a ''real'' plasma of gluons and, eventually, quarks this
interaction is much more complicated. This is true in principle, buth it is
not clear to what consequence: The highly nonlinear nature of the quantum
transport problem might well lead to an enhancement of the effects we observe.
A calculation within another scheme, where the real part of
the self energy function was treated differently \cite{hny96}, 
indeed indicates 
{\em more} quantum oscillations than obtained here.

A final remark of the present paper concerns the use of old-style
classical transport equations for strongly interacting systems 
like colliding nuclei: As we have shown, it is not justified a priori,
i.e., without checking against a solution of the matrix valued 
Schwinger-Dyson equation.
\clearpage
\appendix
\section{Summary of the equations for the spectral function}
In this appendix we present some derivations used above to obtain
$\omega_t$, $\gamma_t$ and $N_{t}$.

The real and imaginary part of the self-energy can be splitted into a
constant part and a part being only proportional to a
$\theta$-distribution in time.
\begin{eqnarray}
{\mbox{Re}}\Sigma^{R/A}_{XP}&=&{\mbox{Re}}\Sigma_i
+\theta(t)({\mbox{Re}}\Sigma_f-{\mbox{Re}}\Sigma_i)=
{\mbox{Re}}\Sigma_i
+\theta(t)\Delta{\mbox{Re}}\Sigma\,,\nonumber\\
{\mathrm i}{\mbox{Im}}\Sigma^{R/A}_{XP}&=&
\mp{\mathrm i}\pi\Gamma_{XP} =
\mp {\mathrm i}\pi \widetilde{g}\gamma_0
\Big(T_i+(T_f-T_i)\theta(t)\Big)=\mp {\mathrm i}\pi \widetilde{g} \gamma_0
\Big(T_i+\Delta T\theta(t)\Big)\,.\nonumber
\end{eqnarray}
The contributions of the constant parts of the self-energy can be
evaluated trivially, when inserting the ansatz for the spectral function.
eq. (\ref{sfan}):
\begin{eqnarray}
&&{\mbox{Re}}\Sigma_i \exp\left(- {\mathrm i}
   \stackrel{\leftrightarrow}{\Diamond}\right) S^{R/A}\nonumber\\
&&= {\mbox{Re}}\Sigma_i \int dE' \frac{{\cal{A}}(E',{\bbox{p}},t)}
{E-E'\rm i\epsilon}
\nonumber\\
&&={\mbox{Re}}\Sigma_i\Big[
   \frac{\omega_t\gamma_0+{\bbox{p}}{\bf{\gamma}}+m_t} 
{2\omega_t(E-\omega_t\pm {\mathrm i} \gamma_t)} 
+ \frac{\omega_t\gamma_0-{\bbox{p}}{\bf{\gamma}}-m_t} 
{2\omega_t(E+\omega_t\pm {\mathrm i} \gamma_t)} \Big]\,,
\;.\end{eqnarray}
More care must be applied in the case of the parts of the
self-energy proportional to the $\theta$-distribution.
At the same time we evaluate the trace over the Dirac matrices.
\begin{eqnarray}
&&\frac{1}{4} \mbox{Tr}\Big[(\Delta{\mbox{Re}}\Sigma\, \theta(t)) 
\exp\left(- {\mathrm i}
   \stackrel{\leftrightarrow}{\Diamond}\right) S^{R/A}\Big]\nonumber\\
&&=\frac{1}{4}\Delta{\mbox{Re}}\Sigma\,  \mbox{Tr} \Big[
 \theta(t)\int dE' \frac{{\cal{A}}(E',{\bbox{p}},t)}{E-E'\pm
   {\mathrm i}\epsilon}
\pm \int dE'\frac{e^{-2{\mathrm i}(E-E')t}\theta(\mp
   t){\cal{A}}(E',{\bbox{p}},t)}
{E-E'\pm {\mathrm i}\epsilon}\Big]
\nonumber\\
&&=\Delta{\mbox{Re}}\Sigma\Big[
\theta(t) \frac{m_t}{2\omega_t}\Big(
\frac{1}{E-\omega_t\pm {\mathrm i}\epsilon}- \frac{1}{E+\omega_t\pm
{\mathrm i}\epsilon}\Big) 
\nonumber\\
&&\qquad
\pm\theta(\mp t) \frac{m_t}{2\omega_t}e^{\pm2\gamma_t t}\Big(
\frac{e^{-2{\mathrm i}(E-\omega_t) t}}{E-\omega_t\pm {\mathrm i}\epsilon}
-\frac{e^{-2{\mathrm i}(E+\omega_t) t}}{E+\omega_t\pm
{\mathrm i}\epsilon}\Big)\Big]\,, 
\end{eqnarray}

\begin{eqnarray}
&&\frac{1}{4} \mbox{Tr}\Big[(\mp {\mathrm i}\pi {\widetilde{g}}\gamma_0\,\Delta T
   \,\theta(t)) \exp\left(- 
   {\mathrm i} 
   \stackrel{\leftrightarrow}{\Diamond}\right) S^{R/A}\Big]\nonumber\\
&&=\frac{1}{4} {\mathrm i}\pi {\widetilde{g}}  \,\Delta T \, \mbox{Tr} \Big[\mp\gamma_0
 \theta(t)\int dE' \frac{{\cal{A}}(E',{\bbox{p}},t)}{E-E'\pm
   {\mathrm i}\epsilon}
- \int dE'\frac{e^{-2{\mathrm i}(E-E')t}\theta(\mp
   t){\cal{A}}(E',{\bbox{p}},t)}
{E-E'\pm {\mathrm i}\epsilon}\Big]
\nonumber\\
&&=\frac{{\mathrm i}\pi {\widetilde{g}}\Delta T}{2}\Big[\mp
\theta(t) \frac{m_t}{2\omega_t}\Big(
\frac{1}{E-\omega_t\pm {\mathrm i}\epsilon}+ \frac{1}{E+\omega_t\pm
   {\mathrm i}\epsilon}\Big) 
\nonumber\\
&&\qquad
-\theta(\mp t) \frac{m_t}{2\omega_t}e^{\pm2\gamma_t t}\Big(
\frac{e^{-2{\mathrm i}(E-\omega_t) t}}{E-\omega_t\pm {\mathrm i}\epsilon}
+\frac{e^{-2{\mathrm i}(E+\omega_t) t}}{E+\omega_t\pm 
{\mathrm i}\epsilon}\Big)\Big]\,,
\end{eqnarray}

\begin{table}[t]
\begin{tabular}{|c||c|c|c|c|c|c|l|}
$\alpha_S(T_c,T)$ &  150   &   175 &  200  &  225  & 250   & 275   & 300  $T$ in MeV
\\ \hline\hline
150        &  0.313 & 0.291 & 0.276 & 0.262 & 0.251 & 0.242 & 0.234
\\ \hline
175        &    --  & 0.313 & 0.294 & 0.279 & 0.267 & 0.256 & 0.248
\\ \hline
200        &    --  &   --  & 0.313 & 0.296  & 0.282 & 0.271 & 0.262
\\ \hline
225        &   --   &   --  &  --   &  0.313  & 0.298  & 0.285  & 0.275     
\\ \hline
250        &   --   &   --  &  --   &   --  & 0.313 & 0.299   & 0.287
\\
$T_c$ in MeV &&&&&&&
\end{tabular}
\caption{The coupling $g$ depending on the critical temperature $T_c$
and the temperature of the quark-gluon plasma $T$.}
\end{table}
\begin{table}[tb]
\begin{tabular}{|c||c|c|c|c|c|c|l|}
$m(T_c,T)$ &  150   &   175 &  200  &  225  & 250   & 275   & 300  $T$ in MeV
\\
in GeV  & & & & & & & 
\\ \hline\hline
150        &  0.121   &  0.137  & 0.152 &  0.167 &  0.181 &  0.196&   0.21
\\ \hline
175        &    --  & 0.142   & 0.157   & 0.172   & 0.187   & 0.202  & 0.216 
\\ \hline
200        &    --  &   --  & 0.162  & 0.177 &  0.192 &  0.207 &  0.222
\\ \hline
225        &   --   &   --  &  --   &  0.182  & 0.197  & 0.212  & 0.228 
\\ \hline
250        &   --   &   --  &  --   &   --  & 0.202   &0.218 &  0.233
\\
$T_c$ in MeV &&&&&&&
\end{tabular}
\caption{The thermal mass $m$ depending on the critical temperature $T_c$
and 
the temperature of the quark-gluon plasma $T$.}
\end{table}
\clearpage
\setlength{\unitlength}{1mm}
\begin{figure}[ht]
\begin{picture}(150,75)
\put(0,0){\includegraphics{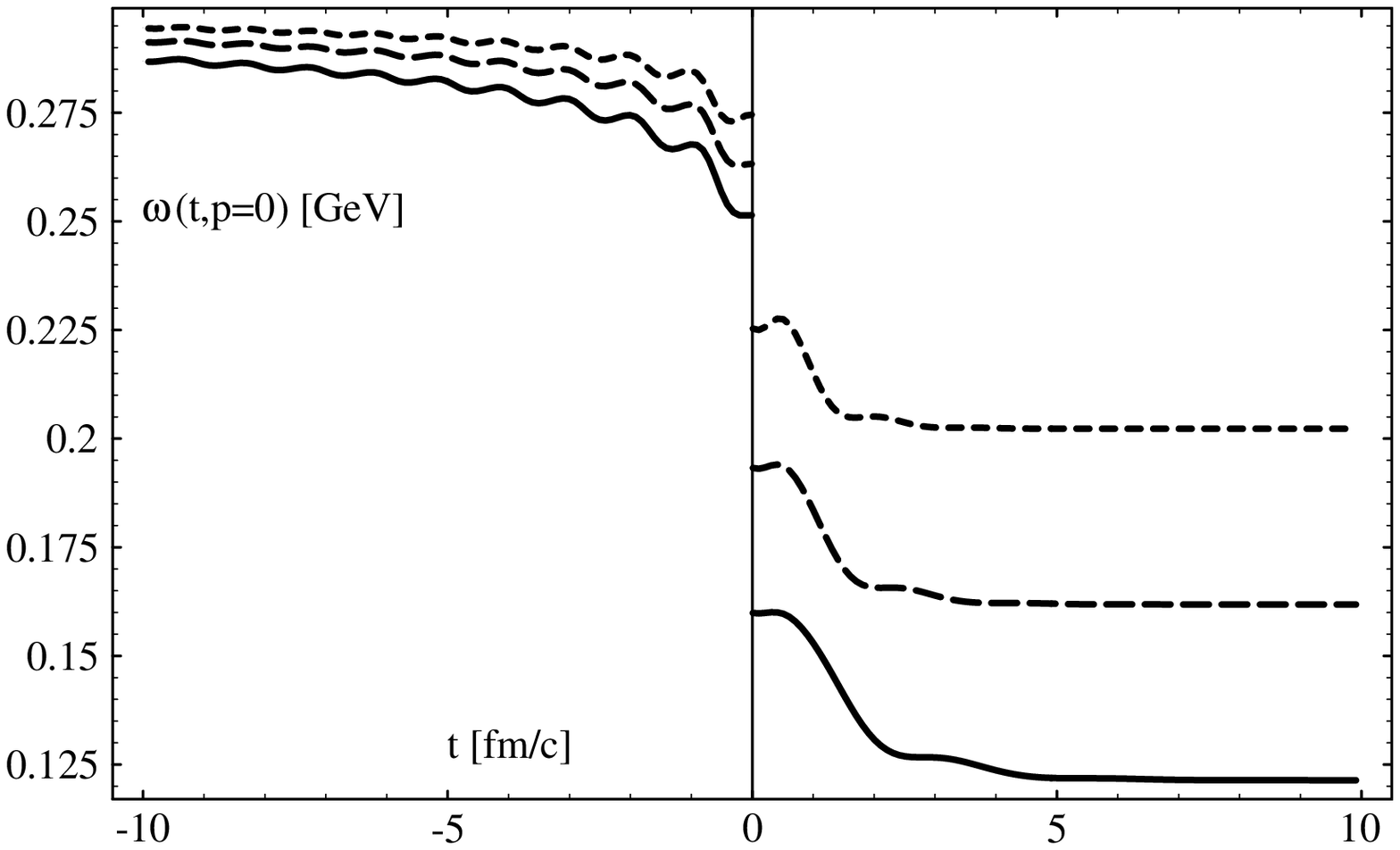}}
\end{picture}
\caption{$\omega_t$ as a function of time,
curve parameter $T_{f}=T_{c}$ with values 250 MeV (short dashed),
200 MeV (long dashed) and 150 MeV (continuous).}
\end{figure}
\begin{figure}[hb]
\begin{picture}(150,75)
\put(0,0){\includegraphics{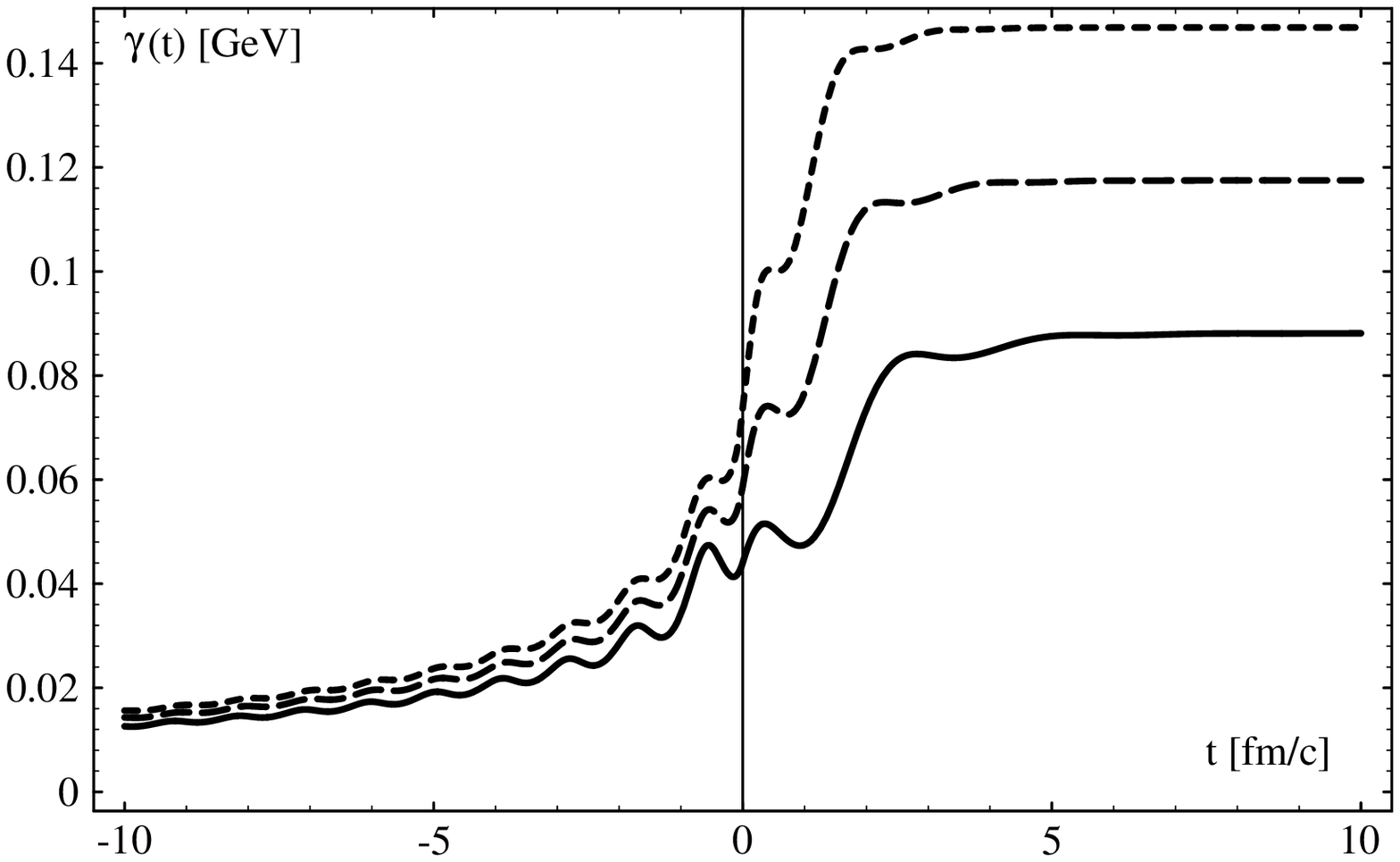}}
\end{picture}
\caption{$\gamma_t$ as a function of time,
curve parameter $T_{f}=T_{c}$ with values 250 MeV (short dashed),
200 MeV (long dashed) and 150 MeV (continuous).}
\end{figure}

\begin{figure}[ht]
\begin{picture}(150,75)
\put(0,0){\includegraphics{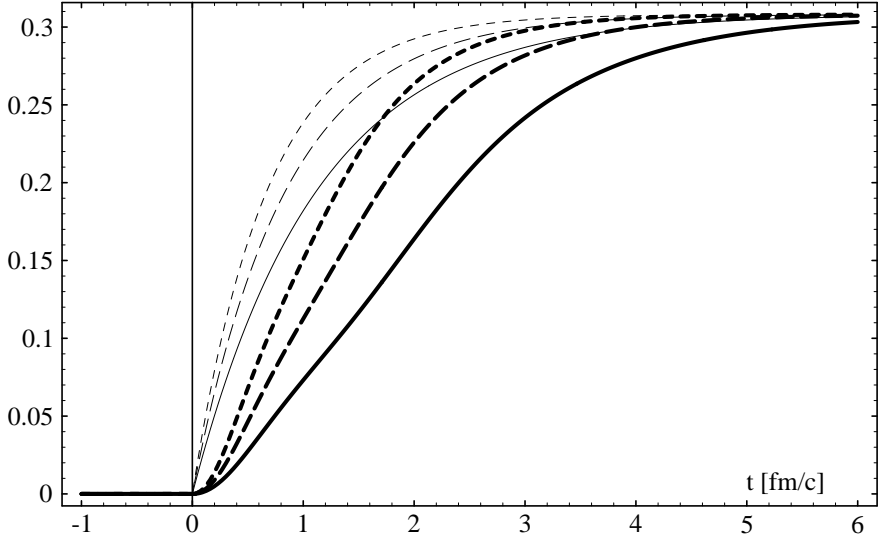}}
\end{picture}
\caption{Classical solution $N^{cl}_{t}$ (thin lines) 
and quantum solution $N_{t}$ (thick lines) of the transport equation,
curve parameter $T_{f}=T_{c}$ with values 250 MeV (short dashed),
200 MeV (long dashed) and 150 MeV (continuous).}
\end{figure}
\begin{figure}[h]
\setlength{\unitlength}{1mm}
\begin{picture}(150,75)
\put(0,0){\includegraphics{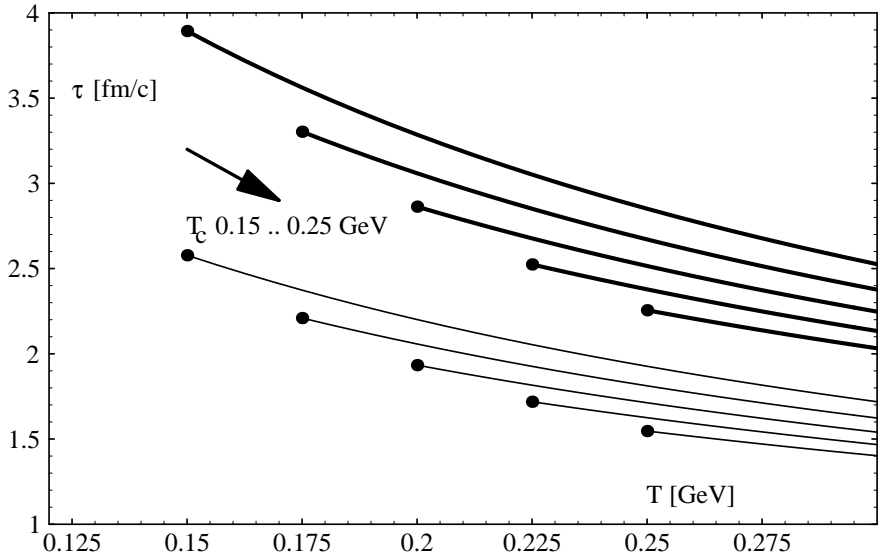}}
\end{picture}
\caption{Relaxation time $\tau$ to reach 90 \% of the final occupation number
for the classical (thin lines) and quantum solution (thick lines).
Curve parameter is the critical temperature $T_{c}=150, 175, \dots 250$ MeV,
the dots indicate the transition point.}
\end{figure}
\end{document}